# Expert Software for the Determination of Juvenile People's Obesity

**Assoc.Prof. Tiberiu Marius Karnyanszky, Ph.D.**
**Assist.Prof. Corina Muşuroi, M.D.**
**Univ.Assist. Carla Amira Karnyanszky**
„Tibiscus" University of Timişoara, Romania

ABSTRACT: Keeping the health condition at the juvenile population is the concern of both the medical staff and the Physical education teachers. Considering the steady tendency of the growth of the number of the youth with excessive weight (overweighed or obese), the intervention to combat this phenomenon must be initiated in the early stage when this condition occurs. The screening method for evaluating the body weight presented in this study uses a calculus program through which, based on the input data (age, sex, weight the width of the skin fold), an evaluation regarding the each juvenile's weight and the body structure can be done. This program can be used in schools, high schools, universities, to signal the weight excess and to appreciate the intervention results for getting a normal weight.

## 1. Overweight and obesity on Earth

Overweight and obesity are defined by World Health Organization (WHO) as abnormal or excessive fat accumulation that may impair health [WHO09].

To determine the obesity state and to classifying overweight and obesity in child or adult populations and individuals, we commonly use the Body mass index (BMI), defined as the weight in kilograms divided by the square of the height in meters ($kg/m^2$).

BMI offers "the most useful population-level measure of overweight and obesity as it is the same for both sexes and for all ages of adults. However, it should be considered as a rough guide because it may not correspond to the same degree of fatness in different individuals" [WHO09].

113

Corresponding to the WHO classifications, "overweight" consists of a BMI equal to or more than 25 and "obesity" consists of a BMI equal to or more than 30. These 25/30 cut-off points provide a benchmark for individual assessment, but there are evidences that risk of chronic disease in populations increases starting to a BMI of 21 [WHO09].

Starting from 2006, the WHO includes BMI standards for infants and young children up to age 5. However, measuring overweight and obesity in children aged 5 to 14 years is challenging because there is not a standard definition of childhood obesity applied worldwide. WHO is currently developing an international growth reference for school-age children and adolescents [WHO09].

WHO's latest projections indicate that globally in 2005:
- approximately 1.6 billion adults (age 15+) were overweight;
- at least 400 million adults were obese;
- at least 20 million children under the age of 5 years are overweight.

WHO also projects that by 2015, approximately 2.3 billion adults will be overweight and more than 700 million will be obese.

## 2. Overweight and obesity in Romania

Romania follows the worldwide tendency: a EUROSTAT processing from 2000 based on national (urban and rural data) examination survey established the percentages of obese/overweight people as presented in figures 1-3.

Particularizing to the juveniles, the concerning of teachers and physicians regarding to the growing and harmonious development of young people established an increasing number of overweight and obese children and teenagers.

The major idea of this pointing out is related to the direct consequences of the obesity, cause of morbidity and mortality. The overweight and the obesity determine deadly diseases as cardiovascular illnesses (myocardial infarction, high blood pressure, attrite), nervous illnesses (paresis, paralysis), nutritional and metabolic illnesses (diabetes). The development of the obesity at early age determine the installation of these diseases in the scholar or teenager time, decreasing the working possibilities, increasing the number of hospitalization, going to social disturbances.

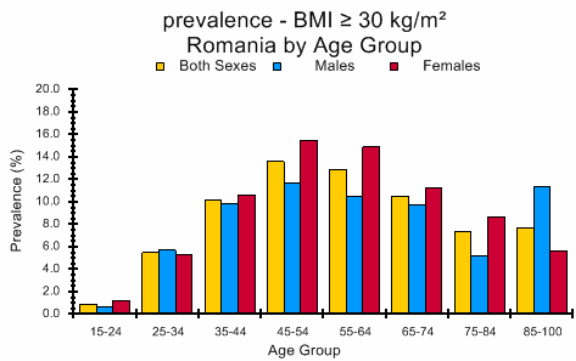

**Figure 1. Obesity state in Romania 2000**

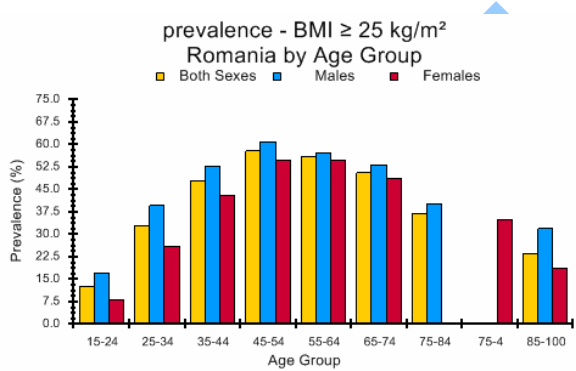

**Figure 2. Overweight state in Romania 2000**

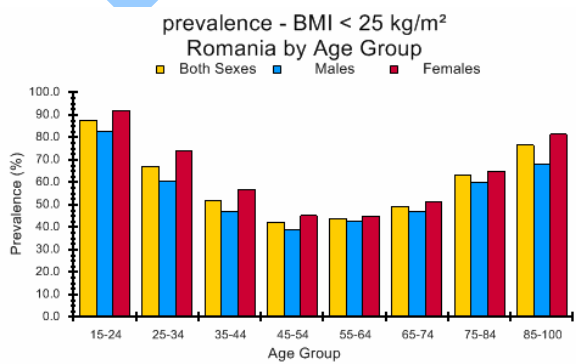

**Figure 3. Normal state in Romania 2000**

The latest "Report on the physic development of the juvenile people in Romania"[1] elaborated by the Institute for Public Health of Bucureşti shows that ([ISP09]):

*"Data centralized from 36 counties with urban population and 23 counties with rural population [...] indicated that [...] 700.617 child were examined regarding to their physic development. A percentage of 74.5% (521734) of them presents a harmonic development and 25.5% (178883) presents a disharmonic development [...]. [...] the percentage of increased disharmonic development is higher that the percentage of decreased disharmonic development."*

## 3. Material and method

This paper proposes a screening method to evaluate the weight of teenagers, at the primary, secondary and high level. This method uses a database to preserve a strictly evidence of all data related to the evolution of youth, and can be used to survey the efficiency of the procedures applied on school or family on the overweight and obese children.

The method uses a computer-based application with an algorithm based on two input parameters: the Body Mass Index (BMI) and the Percent of Adipose Tissue (PAT) related to the Active Body Mass (ABM). The database also contains an ID, the name, the date of birth, the age and sex, the weight, the waist, the dimensions of the fold in the skin.

The weight, tail and folds are measured every evaluation session. On every session the following procedures are performed:
1. The control of corporal weight

It makes a comparison between the measured values and the medium ones (indicated by the Institute of Public Health). For example, Table 1 presents the standard values of body weight for 17 years old boy.

**Table 1. Medium value and standard deviation
of body weight for 17 y.o. boys, urban population**

| Medium | Standard deviation | M-2d | M-d | M+d | M+2d |
|---|---|---|---|---|---|
| 63,473 | 9,035 | 45,403 | 54,438 | 72,508 | 81,543 |

---

[1] http://www.ispb.ro/_stats/visits.php?url=/web/pdf/ro/Sinteze_nationale/ 14_Dezvoltarea_fizica_a_copiilor.pdf&id=1&pagename=FILE:Links/Downloads/pdf/ro/Sinteze_nationale/14_Dezvoltarea_fizica_a_copiilor.pdf

2. <u>The calculation of the BMI</u>
The BMI calculation uses the formula:

$$BMI = \frac{Body\ weight\ (kg)}{Body\ waist^2\ (m^2)} \qquad (1)$$

According to this formula, the teenagers determined with increased BMI are divided into the overweight or obese classes as presented in Table 2:

**Table 2. The International Classification
of underweight, overweight and obesity according to BMI**

| Classification | BMI(kg/m$^2$) | |
| --- | --- | --- |
| | **Principal cut-off points** | **Additional cut-off points** |
| **Underweight** | **<18.50** | **<18.50** |
| Severe thinness | <16.00 | <16.00 |
| Moderate thinness | 16.00 - 16.99 | 16.00 - 16.99 |
| Mild thinness | 17.00 - 18.49 | 17.00 - 18.49 |
| **Normal range** | **18.50 - 24.99** | **18.50 - 22.99** |
| | | **23.00 - 24.99** |
| **Overweight** | **≥25.00** | **≥25.00** |
| Pre-obese | 25.00 - 29.99 | 25.00 - 27.49 |
| | | 27.50 - 29.99 |
| **Obese** | **≥30.00** | **≥30.00** |
| Obese class I | 30.00 - 34-99 | 30.00 - 32.49 |
| | | 32.50 - 34.99 |
| Obese class II | 35.00 - 39.99 | 35.00 - 37.49 |
| | | 37.50 - 39.99 |
| Obese class III | ≥40.00 | ≥40.00 |

**Source:** Adapted from World Health Organization, 1995, 2000 and 2004.

3. <u>The calculation of the PAT and ABM</u>
The weight and BMI determination must be corroborated to the calculation of the ABM, especially if the weight exceeds the standard value. It is well known that sometimes an increased BMI is not related to an excess of adipose, or a normal weight can be related to an excess adipose.

The determination of the body composition can be done using two methods (as presented in [MKK07]):

a. The direct method– the determination of the weight in complete immersion. It's a difficult method, special conditions are required, and so it is rarely used.
b. The indirect method – the determination of the weight by measuring the folds of the skin. It's very accessible so it is often used in practice.

Based on the second method, the determined skin folds are used to compute the body density (BD) by the classic 7-fold-formula with different coefficients for boys and girls:

- **Boys**

$$BD = 1{,}112 - 0{,}00043499 * F + 0{,}00000055 * F^2 - 0{,}00028826 * A \qquad (2)$$

where F is the sum of the 7 folds and A is the boy's age.

- **Girls**

$$BD = 1{,}0970 - 0{,}00046971 * F + 0{,}00000056 * F^2 - 0{,}00012828 * A \qquad (3)$$

where F is the sum of the 7 folds and A is the girl's age.

Now, starting with the BD we can calculate the PAT using the following formulas, depending on age and sex:

| Age | Sex | PAT Formula | |
|---|---|---|---|
| 8-12 | M | $PAT = \dfrac{5.27}{BD} - 4.85$ | (4) |
| 8-12 | F | $PAT = \dfrac{5.27}{BD} - 4.85$ | (5) |
| 13-18 | M | $PAT = \dfrac{5.12}{BD} - 4.69$ | (6) |
| 13-18 | F | $PAT = \dfrac{5.19}{BD} - 4.76$ | (7) |

## 3. The Expert System

This expert system contains a database that uniquely identifies all persons using an ID (the CNP – Personal Numeric Code). Input data are the ID, the age, the waist, the weight, the 7 folds, as shown in figure 4.

Next, the computer makes the determinations and presents the results (BMI, BD, PAT) – figure 5.

All these data are memorized so the user can watch the personal evolution – see figure 6.

**Figure 4. The input data**

**Figure 5. The output data**

**Figure 6. The memorized data**

Based on these memorize data, characterizing the body weight related to the waist, al school level, we can implement a program to assist those overweight or obese child, to comeback to an optimal weight, corresponding to their age and sex.

This action bases on the association between BMI, adipose weight and the distribution of the fat tissue on the body segments, represented by measurements of circumferences. Also, a program of physical exercises can be associated to the excessive adipose areas. A computer-based examination of this data can generate an individual program of training and, further, an analysis of the evolution of the evaluated parameters.

**Conclusions**

This paper presents an operational instrument to monitor the evolution of scholar-age people, based on a computer application. The proposed method can memorize the child's evolution and can be made by the scholar physician or the family doctor. The determination and the admittance of the overweight or obesity state is the first step to a treatment, to a physical adequate behavior, even to a psychological evaluation.